# Theory of Neutron Scattering in the Normal and Superconducting States of $YBa_2Cu_3O_{6+x}$

D. Z. Liu[1], K. Levin[1], and Y. Zha[2]

[1] *James Franck Institute, University of Chicago, Chicago, IL 60637*

[2] *Department of Physics, University of Illinois at Urbana-Champaign, Urbana, IL 61801*

We analyze neutron experiments on $YBa_2Cu_3O_{6+x}$ at various stoichiometries in the superconducting state, within the context of a bi-layer theory which yields good agreement with the normal state Cu-NMR and neutron data as a function of $\omega$, $\mathbf{q}$ and T. A d-wave superconducting state exhibits peaks at $\mathbf{q} = (\pi, \pi, \pi)$ and sharp maxima as a function of $\omega$, at twice the gap frequency. This behavior may have been observed experimentally. The counterpart behavior for other choices of order parameter symmetry is discussed.

In view of the growing experimental support for d-wave superconductivity in the $YBa_2Cu_3O_{6+x}$ cuprate family, it is becoming increasingly important to characterize the magnetic fluctuations in these materials. While a magnetism induced pairing scenario is not uniquely implied by this order parameter symmetry, it is entirely consistent with it. Consequently, neutron experiments with their capability of resolving both the $\mathbf{q}$ and $\omega$ structure of the spin fluctuations are of great interest. They provide the most direct constraints on the normal state spin fluctuations which might drive d-wave superconductivity and in the superconducting state, they contain important signatures of the order parameter symmetry.

In early neutron experiments [1,2], antiferromagnetic peaks ( at $\mathbf{q} = \pi, \pi, \pi$ ) were observed at reduced oxygen stoichiometries . As optimal stoichiometry was approached, the peak widths were found to become relatively broad so that even for $YBa_2Cu_3O_{6.7}$ they corresponded to in plane magnetic correlation lengths which were between 1 and 2 lattice spacings. Nevertheless these peaks were associated with rather low characteristic energy scales ( $\sim$ 20 - 40 meV). Subsequent experiments have reported weak [3,1,4] or absent [5] normal state magnetic peaks for the fully oxygenated material, and, when present, analogously low energy scales. In addition a sharp feature at 41 meV has been noted which may [3] or may not [5] be associated with weak magnetic scattering in the normal state. Of particular interest is the enhancement of this 41 meV feature below $T_c$.

Previous work by our group [6,7] has addressed these normal state data and explained $\mathbf{q}, \omega$ and temperature T dependences as well as trends with variable oxygen stoichiometry, through a combination of Coulomb renormalized bandwidth and weak residual exchange interaction effects. Similar calculations met with the same degree of success for the LaSrCuO family. In contrast to other approaches [8,9] it is argued that at optimal stoichiometry, these materials are not particularly close to magnetic in-stabilities. Incipient localization effects help provide the low energy scales. A three band, large U, ( 1/N ) calculation yields a dynamical susceptibility

$$\chi(\mathbf{q}, \omega) = \frac{\chi^o(\mathbf{q}, \omega)}{1 - J(\mathbf{q})\chi^o(\mathbf{q}, \omega)} \qquad (1)$$

where, unlike the Hubbard RPA theory, the enhancement factor contains additional tight binding $\mathbf{q}$-structure, $J(\mathbf{q}) = -J_o[\cos q_x a + \cos q_y a]$. The Lindhard function $\chi^o$ is given by the usual expression with a three band-Coulomb renormalized energy dispersion [6]. As an independent check on the latter, we have confirmed that the computed Drude-fitted plasma frequencies [6] are in reasonable accord with experiment. Our earliest work emphacized [10] the importance of including realistic Fermi surface shapes in this analysis, which philosophy has been widely adopted by the community. Specifically, for the case of $YBa_2Cu_3O_{6+x}$ we find two Fermi surfaces which are roughly concentric and which are rotated 45° relative to the Fermi surface of the nearest neighbor tight binding model. Both of these effects have been observed experimentally in photoemission ( ARPES ) experiments [11,12]. It should be emphasized that the bandstructure does not suggest there are special nesting vectors (e.g. $(\pi, \pi)$) in YBCO. NMR data, which provide additional constraints on models for $\chi$, have also been reasonably successfully addressed at the Cu site, although the standard hyperfine transfer model for the oxygen relaxation appears to be more problematic.

A bi-layer modulation of the cross section along the c-axis is another important feature of the neutron data. This was found [7] to be a consequence of antiferromagnetic inter- layer correlations $J_\perp$. While an electronic hopping $t_\perp$ will produce a c-axis modulation, in order for the maximum intensity to occur for all frequencies at $q_z = \pi$, this magnetic coupling must be present. To see the effects of $t_\perp$ and $J_\perp$ we re-write Eq(1) as follows:

$$\chi(\mathbf{q}, \omega) = \frac{\chi^o + 2(\chi^o_{22}\chi^o_{11} - \chi^o_{12}\chi^o_{21})(-J_{\perp o} \cos q_z l - J)}{1 - J(\chi^o_{11} + \chi^o_{22}) - (J_{\perp 21}\chi^o_{12} + J_{\perp 12}\chi^o_{21}) + (\chi^o_{11}\chi^o_{22} - \chi^o_{12}\chi^o_{12})(J^2 - J^2_{\perp o})} \,. \tag{2}$$

Here, $\chi^o_{ij}(i, j = 1, 2)$ is the Lindhard function in the layer indices of the bi-layers, and $J(\mathbf{q}) = -J_o(\cos q_x + \cos q_y)$ is the antiferromagnetic superexchange within the plane. The antiferromagnetic interlayer coupling $J_\perp(\mathbf{q})$ is defined as $[J_{\perp 12}(\mathbf{q}) = -J_{\perp o}e^{iq_z l} = J^*_{\perp 21}]$ which is included in the usual RPA procedure, $l$ is the distance between the closest $CuO_2$ layers in $YBa_2Cu_3O_{6+x}$. The electronic coupling $t_\perp$ is embedded in the band structure which enters $\chi^o$:

$$\chi^o = \chi^o_{11} + \chi^o_{12} + \chi^o_{21} + \chi^o_{22} \tag{3}$$
$$= \frac{1 + \cos q_z l}{2}(\chi^o_{++} + \chi^o_{--}) + \frac{1 - \cos q_z l}{2}(\chi^o_{+-} + \chi^o_{-+})$$

where $\chi^o_{\alpha,\beta}(\alpha, \beta = +, -)$ are the Lindhard functions in the bi-layer bonding and antibonding band indices, and the transformation between the layer- and band- representations is:

$$\chi^o_{11} = \chi^o_{22} = \frac{1}{4}[\chi^o_{++} + \chi^o_{--} + \chi^o_{+-} + \chi^o_{-+}] \tag{4}$$
$$\chi^o_{12} = \chi^{o*}_{21} = \frac{e^{iq_z l}}{4}[\chi^o_{++} + \chi^o_{--} - \chi^o_{+-} - \chi^o_{-+}]$$

Given a satisfactory theory of the normal state data, the implications for a $d$-wave superconducting state can be readily derived. We previously predicted [6] a low frequency ($< 10$ meV) signature associated with a commensurate to incommensurate transition below $T_c$. This $d$-wave signature has not yet been observed [13,14]. In view of the recent experimental focus on higher frequencies, at 41 meV, it is important to extend theoretical calculations to this regime.

In our numerical analysis we use the same intra-band parameters for $YBa_2Cu_3O_7$ and $YBa_2Cu_3O_{6.7}$ as were used earlier [6] to address NMR and neutron data. The effects of the superconductivity enter Eq(2) in the usual way [6] via the Lindhard function. Figure 1 shows the wave-vector dependence in $YBa_2Cu_3O_7$ of the imaginary part of $\chi$ for the $d$-wave case and in the inset, for the normal metal. In addition we have taken $t_\perp = 40 meV$ and $J_{\perp o} = 10 meV$. Here the frequency $\omega = 41 meV$ is slightly larger than twice the $d$-wave gap frequency $2\Delta_o$. Following our solution of the bi-layer gap equation, Ref [15], we take the two in- phase, $d$-wave gaps in the bi-layer bands to be identical [16].

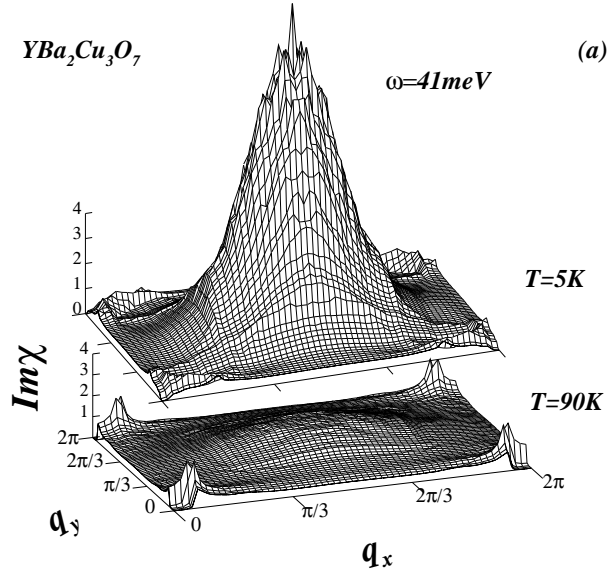

FIG. 1. Calculated neutron scattering form factor $Im\chi$ as a function of $(q_x, q_y)$ for $YBa_2Cu_3O_7$. Here it is assumed that the superconducting gap $\Delta_o = 20 meV$.

The results of these calculations are in reasonable agreement with Ref [5], although the calculated width is somewhat wider than observed. The dependence on $\mathbf{q} = (q_x, q_y)$ is similar for both $q_z = 0$ and $q_z = \pi$, but the amplitude is only significant for latter wave vector. This is a consequence of the perpendicular exchange interaction $J_\perp$. The mechanism for the formation of the strong peak below $T_c$, is as conjectured in Ref. [5], a consequence of pair creation with one electron in the + lobe and the other electron in the - lobe. However, we find that the underlying fermiology of the system also plays a role. The separation of the two lobes is near $(q_x, q_y) = (\pi, \pi)$. This enhances the weight around $(q_x, q_y) = (\pi, \pi)$ in the Lindhard function. The further inclusion of an antiferromagnetic exchange $J(\mathbf{q})$ creates a strong peak at $(\pi, \pi)$. The position of the Fermi surface nesting vectors can displace this maximum away from $(\pi, \pi)$. Thus in LaSrCuO (where there is strong nesting close to but not at $(\pi, \pi)$ there is a general enhancement of the $(q_x, q_y) = (\pi, \pi)$ cross section at twice the gap frequency, but the global maxima appear at the incommensurate wave-vectors of the normal state neutron cross section. We estimate that the counterpart of these effects should be seen in optimally doped LaSrCuO at frequencies of roughly $2\Delta_o \geq 14 meV$.

In Figures 2 are plotted the frequency dependences of the $(\pi, \pi)$ feature (solid line), with the normal state result shown as the dashed line for $YBa_2Cu_3O_7$ (a) and $YBa_2Cu_3O_{6.7}$ (b). The normal state has a weak maximum in the de- oxygenated case and an even smaller feature at optimal stoichiometry, both of which are associated with their respective Van Hove energies. In the absence of antiferromagnetic interaction effects, the width of the superconducting frequency peak is determined by $2\Delta_o$ at the lower end and the Van Hove energy at the upper. In the presence of $J(\mathbf{q})$, both energies are softened by spin fluctuation effects. The gap feature rises more sharply the larger is $J(q)$. For the parameters that we have chosen from numerous other considerations [6]this gives rise to a peak width which is considerably sharper in the superconducting state, but which should be close to the resolution limited value found experimentally. As the oxygen concentration is reduced, these two energy scales pull apart and the peak broadens somewhat. Ultimately, two distinct frequency features appear, as may have been observed by Bourges *et al* [4] [17]. The presence of two bands leads to the two rather weak Van Hove features of the figure. It should be stressed that the frequency regime over which the normal and superconducting contributions occur is rather similar. This is a consequence of narrow band effects. If the band-widths were taken to be broader, the (weak) normal feature will be at higher frequencies relative to that of the superconducting state. Thus, in this system, also, fermiology plays a role.

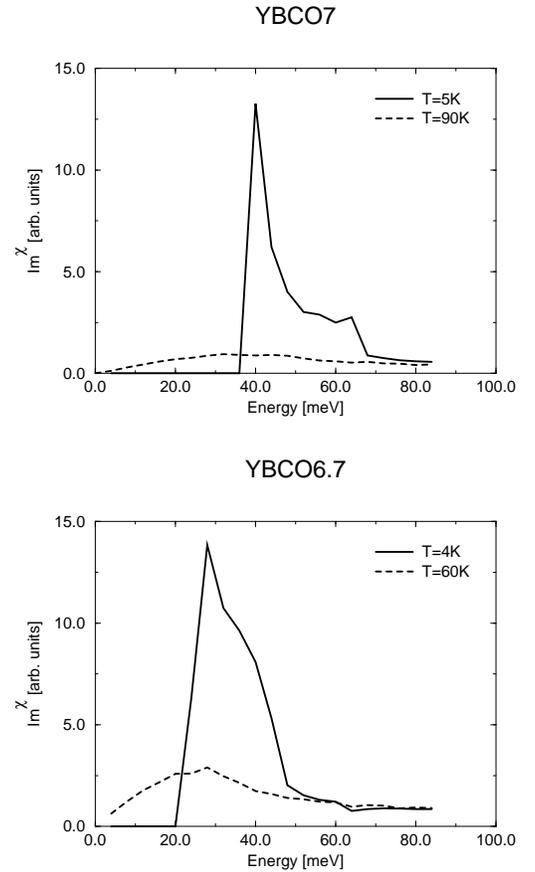

FIG. 2. Calculated frequency dependence of Im$\chi$ for (a) $YBa_2Cu_3O_7$ with same parameters as in Fig.1; and (b) predictions for $YBa_2Cu_3O_{6.7}$ by assuming $\Delta_o = 15 meV$.

Finally in Figure 3, the temperature dependence of the peak height at fixed $\omega = 41 meV$ is indicated for the fully oxygenated sytem. This shows the overall scale of the superconducting feature relative to the normal state background and its rather dramatic temperature onset.

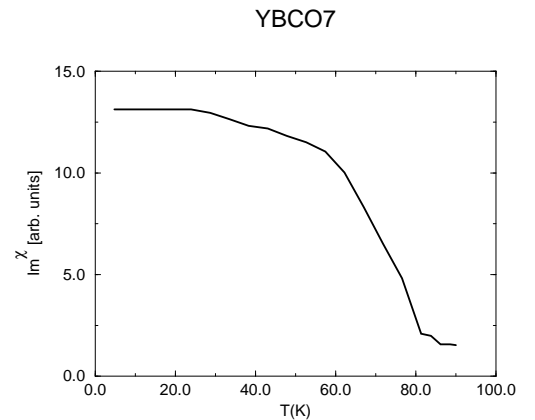

FIG. 3. Calculated temperature dependence of Im$\chi$ for $YBa_2Cu_3O_7$ with same parameters as in Fig.1.

We have performed similar studies for an alternate candidate for the order parameter symmetry. As shown in Ref [15], when the $c$- axis modulation of the spin fluctuation spectra are included in the gap equation, the two in phase $d$-states of the bi-layer unit, compete with two out of phase $s$-states. In the presence of weak orthorhombicity, ( and for particular parameter choices) this state exhibits the features of a $d_{x^2-y^2}$ state, having one gap elongated along the $a$ and the other along the $b$ axis of the crystal. In this way results compatible with Josephson experiments on YBaCuO [18] can be obtained in a bi-layer $s$-wave system. Moreover, for this state there will be enhancements of the superconducting cross section below $T_c$ along the $(\pi, \pi)$ direction, which are rather similar to those of the in phase d-wave case discussed above.

In conclusion, we have applied a theory which has been previously extensively compared to normal state magnetic data, in order to ascertain the neutron signatures of d-wave pairing. While at low frequencies incommensurate peaks are expected, at higher $\omega$ commensurate structure will occur associated with a narrow frequency peak. The latter feature may have been observed experimentally by a number of groups. In the present scenario, the sharp onset of the frequency peak indicates that there is some residual exchange interaction present near optimal stoichiometry. On the otherhand, the rather broad $\mathbf{q}$-space features observed in the normal state for both $YBa_2Cu_3O_7$ and $YBa_2Cu_3O_{6.7}$ suggests that the magnetism is relatively weak for the highest transition temperature systems. While this analysis shows consistency with $d$-wave pairing, an alternate orthorhombic bi-layer model with two out of phase $s$- wave bandgaps cannot be ruled out. In order to fully verify the origin of the 41 meV feature, analogous effects should be searched for in the reduced oxygen and LaSrCuO systems. Our calculations should provide some guidelines for future experiments on these materials.

We acknowledge stimulating conversations with B. Keimer. This work is supported by NSF-DMR-91-20000 and NSF-DMR-94-16926 through the Science and Technology Center for Superconductivity.